\documentclass[%
reprint,
showkeys,
superscriptaddress,
preprintnumbers,
nofootinbib,
amsmath,amssymb,
aps,
prd,
floatfix,
]{revtex4-2}
\usepackage{afterpage}
\usepackage{subfigure}
\usepackage{rotating} 
\usepackage{booktabs}
\usepackage{graphicx}
\usepackage{dcolumn}
\usepackage{bm}
\usepackage{amsmath}
\usepackage{natbib}
\usepackage{hyperref}

\usepackage{xcolor}

\hypersetup{
    colorlinks = true,
    allcolors = blue,
}

\newcommand{\myfig}[1]{Fig.~\ref{#1}}
\newcommand{\mytab}[1]{Table~\ref{#1}}

\begin{document}

\preprint{Song, et al., \href{https://doi.org/10.1103/mcf5-1wvt}{Phys.Rev.D 113, 064028 (2026)} }

\title{Gravitational-wave constraints on noncommutative spacetime from GW190814}

\author{Hanlin Song}
\affiliation{Leicester International Institute, Dalian University of Technology, Panjin 124221, China}
\affiliation{School of Physics, Peking University, Beijing 100871, China}

\author{Hao Li}
\email{haolee@cqu.edu.cn}
\affiliation{Department of Physics, Chongqing University, Chongqing 401331, People's Republic of China}

\author{Zhenwei Lyu}
\email{zwlyu@dlut.edu.cn}
\affiliation{Leicester International Institute, Dalian University of Technology, Panjin 124221, China}

\author{Jie Zhu}
\email{jiezhu@cqu.edu.cn}
\affiliation{Department of Physics, Chongqing University, Chongqing 401331, People's Republic of China}

\author{Jun-Chen Wang}
\affiliation{School of Physics, Peking University, Beijing 100871, China}

\author{Peixiang Ji}
\affiliation{School of Physics, Peking University, Beijing 100871, China}

\begin{abstract}
Recent advances in noncommutative geometry and string theory have stimulated increasing research on noncommutative gravity. The detection of gravitational waves~(GW) opens a new window for testing this theory using observed data. In particular, the leading correction from noncommutative gravity to the GW of compact binary coalescences appears at the second post-Newtonian~(2PN) order.  This correction is proportional to the dimensionless parameter $\Lambda\equiv|\theta^{0i}|/(l_Pt_P)$, where $\theta^{0i}$ denotes the antisymmetric tensor characterizing noncommutative spacetime, and $l_P, t_P$ represent the Plank length and time, respectively. Previous study have used the phase deviation from general relativity at the 2PN order, as measured in GW150914, to constrain noncommutative gravity, resulting in an upper bound of $\sqrt{\Lambda}\lesssim3.5$. Another analysis, based on multiple events from the GWTC-1 catalog, has obtained consistent bounds.  In this work, we construct the noncommutative gravity waveform in the Parameterized Post-Einsteinian framework. Based on the \texttt{IMRPhenomXHM} template, we incorporate both the dominant (2,2) mode and several higher-order modes, including (2,1), (3,3), (3,2), and (4,4). We first reanalyze the GW150914 with a Bayesian parameter estimation and derive a 95th percentile upper bound on noncommutative gravity, obtaining $\sqrt{\Lambda}<0.68$.  We then analyze GW190814 and obtain an even tighter 95th percentile upper bound of $\sqrt{\Lambda}<0.46$, which corresponds to a characteristic noncommutative gravity energy scale above
$2.2\,E_P$ or a length scale below $0.46\,l_P$. This represent the strongest constraint on noncommutative gravity derived from real GW observations to date.
\end{abstract}

\maketitle

\section{Introduction}
Gravitational waves~(GW) stand as one of the cornerstone predictions of Einstein's general relativity~(GR).
Since the first detection of the GW from a binary black hole~(BBH) coalescence on September 14, 2015, by the LIGO Scientific Collaboration and Virgo Collaboration~\cite{LIGOScientific:2016aoc}, over 300 GW events have been observed by the network of LIGO, Virgo and KAGRA GW observatories during the O1, O2, and O3 observation runs, as well as the first 23 months of O4~\cite{LIGO2025}. These GW events originate from different types of compact binary coalescence~(CBC), including the BBHs, binary neutron stars~(BNSs) and neutron star-black hole binaries~(NSBHs). With the open science data provided by the Gravitational-Wave Transient Catalogs~(GWTC), such as GWTC-1~\cite{LIGOScientific:2018mvr}, GWTC-2~\cite{LIGOScientific:2020ibl}, GWTC-2.1~\cite{LIGOScientific:2021usb}, GWTC-3~\cite{KAGRA:2021vkt}, and the upcoming  GWTC-4, these valuable observations offer an unparalleled  opportunity to test general relativity and alternative theories of gravity~\cite{LIGOScientific:2019fpa, LIGOScientific:2020tif, LIGOScientific:2021sio, Yunes:2025xwp}.

The Einstein's theory of GR is one of the pillars of modern physics and has passed all experimental tests to date with remarkable precision~\cite{Ghez:2003qj, Psaltis:2008bb, Will:2014kxa, Freire:2024adf}. These tests include observations from the Solar System, binary pulsar, stellar orbits around the Galactic Center, and cosmological measurements. Notably, most of the aforementioned tests are performed in quasistationary and quasilinear weak-field regimes. The detection of GWs opens a new window for testing GR in highly dynamical and strong-field conditions (see, e.g., Ref.~\cite{Yunes:2025xwp} as a recent review). GR can be tested in a theory-independent manner, without relying on any specific alternative theory, by performing consistency tests between observed GW data and GR predictions for events in GWTC-1, GWTC-2, GWTC-2.1, and GWTC-3~\cite{Arun:2006yw, Arun:2006hn, Mishra:2010tp, LIGOScientific:2019fpa, LIGOScientific:2020tif, LIGOScientific:2021sio, Yunes:2025xwp}. These tests include the residual test, inspiral-merger-ringdown consistency test, phenomenological parameter deviation test, and the GW propagation test. For example, detailed consistency test of GR have been performed with GW150914~\cite{LIGOScientific:2016lio} and GW170817~\cite{LIGOScientific:2017zic}, and no evidence for deviations from GR have been founded. Another strategy for testing GR is to directly confront observed data with waveforms derived from alternative theories of gravity. This approach enables non-GR parameters to be constrained in a model-dependent manner. To facilitate such analyses, the \textit{parameterized post-Einsteinian}~(ppE) formalism was proposed by Yunes and Pretorius~\cite{Yunes:2009ke}, providing a generic way to capture the non-GR effects. The non-GR waveform is constructed  by parameterizing the modifications introduced by a specific theory as corrections to the phase and amplitude of the GR tensorial polarizations. Within this manner, several alternative gravity theories have been studied and tested,  including the scalar-tensor theory~\cite{Yagi:2009zm, Yagi:2009zz, Freire:2012mg,Chatziioannou:2012rf, Wex:2014nva, Yunes:2016jcc, Sennett:2016klh, Zhang:2017sym, Gong:2017kim, Shao:2017gwu, Zhao:2019suc, Niu:2021nic, Xu:2021kfh, Takeda:2023wqn, Tan:2023fyl, Chen:2024pcn, Xie:2024xex, Ji:2024gdc}, Einstein-dilaton Gauss-Bonnet Gravity~\cite{Yagi:2011xp,Yagi:2012gp,LIGOScientific:2016vlm, Yunes:2016jcc,DeLaurentis:2016jfs, LIGOScientific:2017vox,Nair:2019iur,Yamada:2019zrb,Tahura:2019dgr, Chatziioannou:2019dsz, Carson:2020ter, Lyu:2022gdr, Minamitsuji:2022tze, Julie:2024fwy, Sanger:2024axs, Gao:2024rel,Wang:2023wgv, Luo:2024vls}, dynamical Chern-Simons gravity~\cite{ Yagi:2012vf, Yagi:2012ya,Yunes:2016jcc,Li:2022grj,Li:2023lqz}, Einstein-{\AE}ther theory~\cite{Jacobson:2007veq, Hansen:2014ewa, EmirGumrukcuoglu:2017cfa, Chamberlain:2017fjl, Gong:2018cgj}, Khronometric theory~\cite{Hansen:2014ewa, Chamberlain:2017fjl}, varying-G theory~\cite{Chamberlain:2017fjl, Tahura:2018zuq, Tahura:2019dgr,Vijaykumar:2020nzc,  Barbieri:2022zge,Wang:2022yxb, An:2023rqz, Sun:2023bvy,  Yuan:2024duo}, as well as the noncommutative gravity~\cite{Chamseddine:2000si,Aschieri:2005yw,Calmet:2005qm,Aschieri:2005zs,Calmet:2006iz,Mukherjee:2006nd,Szabo:2006wx,Kobakhidze:2006kb}. A more comprehensive review can be found in the work of Tahura and Yagi~\cite{Tahura:2018zuq}.  

Noncommutative spacetime\footnote{Details of the theory are introduced in Sec.~\ref{theory}, and more information can be found in the references therein.} has been tested utilizing various methods, including low-energy precision experiments~\cite{Mocioiu:2000ip,Chaichian:2000si}, Lorentz invariance violation~\cite{Carroll:2001ws}, as well as cosmological observations~\cite{Joby:2014oee,Calmet:2015fma}, and the scale of noncommutativity is argued to be no larger than the inverse TeV scale~\cite{Calmet:2004dn}.
Besides, with the observation of GW events, the noncommutative gravity has also been tested in several studies~\cite{Kobakhidze:2016cqh, Jenks:2020gbt}. For example, Kobakhidze  {\it et al.}~\cite{Kobakhidze:2016cqh} used the phase deviation from GR at the 2PN order, as measured in GW150914, obtaining an projected upper bound of $\sqrt{\Lambda}\lesssim3.5$, where \(\Lambda=|\theta^{0i}|/l_P t_P\)\footnote{The Planck length \(l_P=\sqrt{\hbar G/c^3}\simeq1.6\times 10^{-35}~\text{m}\) and the Planck time \(t_P=\sqrt{\hbar G/c^5}\simeq 5.4\times 10^{-44}~\text{s}\).}, and \(\theta^{\mu\nu}\) is the antisymmetric tensor of noncommutativity to be explained later.
Jenks {\it et al.}~\cite{Jenks:2020gbt} analyzed several events from GWTC-1 including GW151225, GW170608, GW170814, and GW170817, deriving bounds consistent with the aforementioned result. The observational data from the pulsar system PSR~J0737-3039A/B were also analyzed by Jenks {\it et al.}~\cite{Jenks:2020gbt}, although the derived bounds are weaker than those obtained from GW observations. Recent studies~\cite{Perkins:2020tra, Huang:2024ylf} further investigated the potential of future GW observatories to probe noncommutative gravity, such as Cosmic Explorer~\cite{Reitze:2019iox}, Einstein telescope~\cite{Punturo:2010zz}, LISA~\cite{LISA:2017pwj}, Taiji~\cite{Hu:2017mde, Du:2025xdq} and TianQin~\cite{TianQin:2015yph}. 

In this work, we focus on testing the noncommutative gravity in the ppE framework. We analyze the observed GW data of GW150914 and GW190814 from LIGO Hanford and LIGO Livingston, which are generated from the coalescence of a BBH system with component masses of \(36~\text{M}_\odot\) and \(29~\text{M}_\odot\), and the coalescence of a $23~\rm  M_{\odot}$ and a $2.6~\rm  M_{\odot}$ compact object, respectively~\cite{LIGOScientific:2020zkf}.  We construct the non-GR waveform within the ppE framework based on the \texttt{IMRPhenomXHM} template at the inspiral stage, both the dominant (2,2) mode and several higher-order modes are considered, including (2,1), (3,3), (3,2), and (4,4). We also consider the merger and ringdown phase with $C^0$ correction~\cite{Bonilla:2022dyt}.  At first, we reanalyze the GW150914 and derive a 95th percentile upper bound of $\sqrt{\Lambda}<0.68$. We then analyze the GW190814, which features the most extreme symmetric mass ratio among events in GWTC-1, GWTC-2, GWTC-2.1, and GWTC-3~\cite{GWOSC_GWTC}, along with a relatively high signal-to-noise ratio (SNR). The results yield a tighter 95th percentile upper bound of $\sqrt{\Lambda}<0.46$.
This is the strongest constraint on noncommutative gravity derived from GW detection to date.

This paper is organized as follows. We first review theoretical development of noncommutative gravity and the derived ppE waveform in previous works in Sec.~\ref{nonconmmugrav}. We then present our data analysis procedures in detail in Sec.~\ref{data_analysis}. In Sec.~\ref{sec:results}, we discuss and conclude our results on constraining noncommutative gravity with GW150914 and GW190814. Geometrical units $G=\hbar=1$ are adopted throughout this paper, while we keep the speed of light \(c\) for later convenience.

\begin{figure*}[ht] 
	\centering 
    \begin{minipage}{0.46\textwidth}
        \centering
        \includegraphics[width=0.95\linewidth]{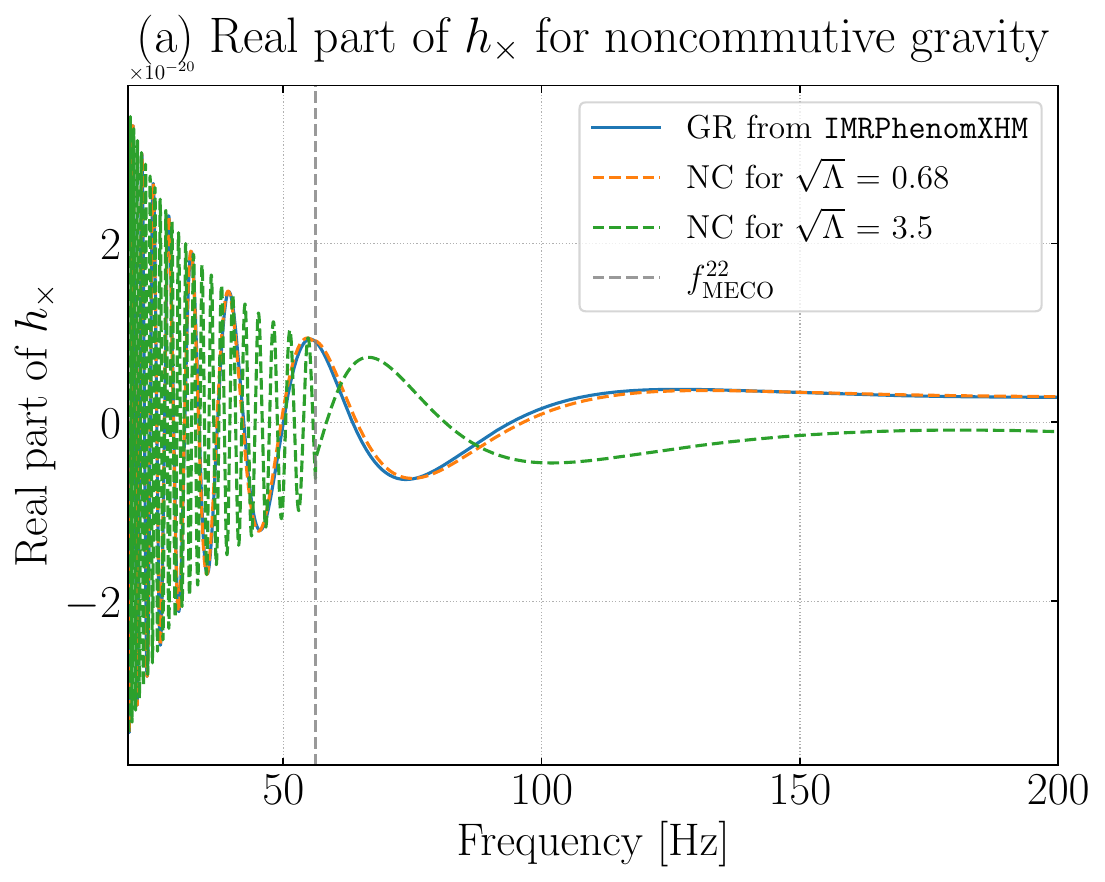}
    \end{minipage}\hfill
    \begin{minipage}{0.49\textwidth}
        \centering
        \includegraphics[width=0.90\linewidth]{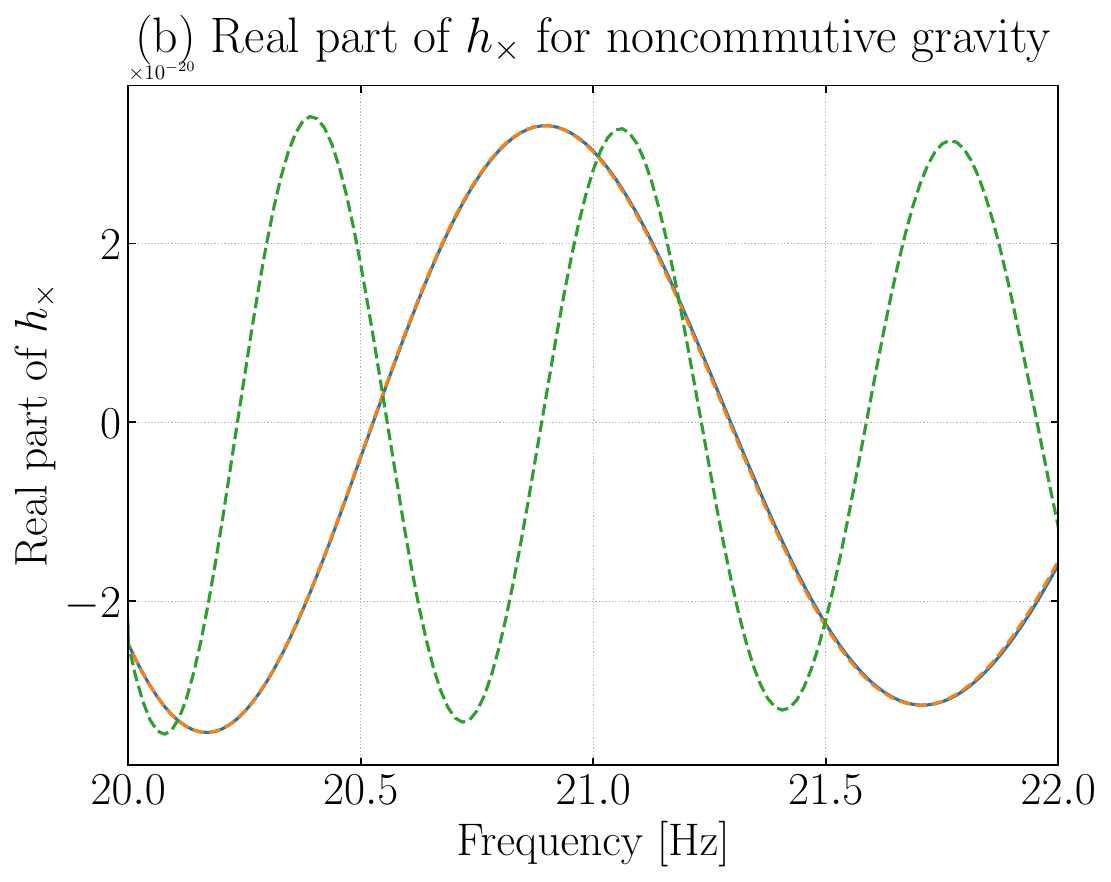} 
    \end{minipage}
    \caption{The waveform of the real part of $h_\times$ for a GW150914-like event is shown for general relativity and noncommutative gravity. The left panel displays the GR waveform, the noncommutative gravity waveform with $\sqrt{\Lambda} = 3.5$ obtained in Ref.~\cite{Kobakhidze:2016cqh}, and the waveform corresponding to the 95th percentile upper bound on $\sqrt{\Lambda}=0.68$ for GW150914 derived in this work. The dashed vertical line denotes the MECO frequency of 22 mode. The right panel shows the waveform in the frequency range from 20 Hz to 22 Hz, with both the low-frequency cutoff and the reference frequency set to 20 Hz.  
    \label{waveforms}}
\end{figure*}

\section{Noncommutative gravity\label{nonconmmugrav}}
In this section, we first present the theoretical foundations of noncommutative spacetime and gravity. We then review noncommutative corrections to the orbital motions and GW waveforms of two inspiraling black holes~(BHs) up to 2PN order. These corrections are subsequently mapped into the ppE framework to enable efficient data analysis.

\subsection{Theory\label{theory}}

Noncommutative spacetime was first proposed by W. Heisenberg to circumvent ultraviolet divergences in quantum field theories. However, following its initial realization~\cite{Snyder:1947nq}, the concept attracted limited interest until revolutionary advances in noncommutative geometry~\cite{connesNoncommutativeDifferentialGeometry1985,Connes:1994yd} and—most significantly—the natural emergence of quantized spacetime in string theory under specific conditions~\cite{Ardalan:1998ce,Seiberg:1999vs} (see also reviews~\cite{Douglas:2001ba,Szabo:2001kg}). In this work, we focus on the string-theoretic approach to noncommutative spacetime, promoting spacetime coordinates to operators satisfying~\footnote{We only consider the simplest theoretical set-up where \([\hat p_\mu,\,\hat p_\mu]=0\), \([\hat x^\mu,\, \hat p_\nu]=i\delta^\mu_\nu\) and \(\theta\) invertible, but generalizations are also possible~(see~\cite{Douglas:2001ba,Szabo:2001kg} for example).}
\begin{equation}
    [\hat x^\mu,\,\hat x^\nu]=i\theta^{\mu\nu},\label{noncommutativerelation}
\end{equation}
where \(\theta\) is a constant antisymmetric matrix whose components are real \(c\)-numbers. Consequently, Eq.~\eqref{noncommutativerelation} implies an uncertainty relation \(|\Delta \hat x^i|\ge |\theta^{0i}|/2\), analogous to the quantum mechanical uncertainty principle. Thus, \(\theta^{\mu\nu}\) quantifies the fundamental fuzziness of spacetime.

In general, noncommutative effects modify both the Standard Model of particle physics~(SM) and the theory of gravity. However, unlike the noncommutative extension of the SM, noncommutative gravity theories can be formulated through diverse approaches yielding distinct theoretical outputs---all emerging at order \(\theta^2\), though.
The model dependence and high complexity~\cite{Chamseddine:2000si,Aschieri:2005yw,Calmet:2005qm,Aschieri:2005zs,Calmet:2006iz,Mukherjee:2006nd,Szabo:2006wx,Kobakhidze:2006kb} of these corrections make constraining spacetime fuzziness via GWs challenging, as no GW solution currently incorporates \(\theta^2\) corrections to the Hilbert-Einstein action.
Consequently, Kobakhidze and collaborators~\cite{Kobakhidze:2016cqh} argue that---based on the observation that all noncommutative corrections to spacetime geometry begin at \(\theta^2\) order~\cite{Calmet:2006iz,Mukherjee:2006nd}---noncommutative corrections to gravity itself may be neglected. Only noncommutative effects on classical matter sources in GW generation need be considered.

Following Kobakhidze \textit{et al.}~\cite{Kobakhidze:2007jn,Kobakhidze:2016cqh}, the total energy-momentum tensor for a binary system is approximated as
\begin{equation}
    T^{\mu\nu}_{\text{NC}} = T^{\mu\nu}_{\text{GR}} + T^{\mu\nu}_{\text{NC},\theta^2},\label{emt1}
\end{equation}
where \(T^{\mu\nu}_{\text{GR}}\) denotes the standard energy-momentum tensor for two point masses~\cite{Blanchet:2000cw,Kobakhidze:2016cqh}, and the \(\theta^2\)-order noncommutative correction reads~\cite{Kobakhidze:2016cqh}:
\begin{align}
    T^{\mu\nu}_{\text{NC},\theta^2}(\bm{x},t)=&\frac{m_1^3\Lambda^2}{8c^4}v_1^\mu(t)v_1^\nu(t)\theta^k\theta^l\partial_{k}\partial_{l}\delta^3\big(\bm{x}-\bm{y}_1(t)\big)\nonumber \\ &+(1\leftrightarrow2),\label{emtnc1}
\end{align}
with the subscript ``$\mathrm{NC}$" denotes noncommutative gravity, \(m_i\) (\(i=1,2\)) being the component masses, \(\bm{y}_i(t)\) their trajectories, and \(v_i^\mu(t) = \left( c, d\bm{y}_i/{dt} \right)\) the four-velocities. The parameters \(\Lambda\) and \(\theta^i\) satisfy \(\Lambda \theta^i \equiv \theta^{0i} / (l_P t_P)\), where \(\theta^i\) is a dimensionless unit vector. The parameter \(\Lambda\) represents the primary constraint target of this work.
Some remarks on Eq.~\eqref{emtnc1} are warranted:
\begin{itemize}
    \item It represents a 2PN-order correction, exclusively accounting for noncommutative effects at this order.
    \item Quantum effects on the energy-momentum tensor are neglected, as we consider only astrophysical objects with masses $m > M_\odot$~\cite{Kobakhidze:2016cqh}.
\end{itemize}
Using the modified energy-momentum tensor~\eqref{emtnc1} and assuming no corrections to the gravity sector, the modified orbital separation $r$ and the GW frequency evolution $\dot f$ are given by~\cite{Kobakhidze:2016cqh,Tahura:2018zuq}:
\begin{align}
    r &= r_{\text{GR}} \left(1 + \frac{1}{8} \eta^{-4/5} (2\eta - 1) \Lambda^2 u^4 \right), \label{orbitalseparation} \\
    \dot{f} &= \dot{f}_{\text{GR}} \left(1 + \frac{5}{4} \eta^{-4/5} (2\eta - 1) \Lambda^2 u^4 \right), \label{frequencyevolution}
\end{align}
respectively. Following the conventions of Tahura and Yagi~\cite{Tahura:2018zuq}, we define key quantities as follows:
\begin{align}
    &\text{Total mass: }  M = m_1 + m_2, \nonumber \\
    &\text{Mass ratio: }  q = m_1 / m_2 \geq 1, \nonumber \\
    &\text{Symmetric mass ratio: }  \eta = \frac{m_1 m_2}{M^2}, \nonumber \\
    &\text{Chirp mass: }  \mathcal{M}_c = \frac{(m_1 m_2)^{3/5}}{M^{1/5}}, \nonumber \\
    &\text{Orbital angular frequency: } \Omega = \pi f, \label{definition}
\end{align}
with leading order auxiliary relations~\cite{Cutler:1994ys,Blanchet:1995ez,Tahura:2018zuq}
\begin{equation}
    r_{\text{GR}} = (M / \Omega^2)^{1/3}, \quad 
    \dot{f}_{\text{GR}} = \frac{96}{5\pi \mathcal{M}^2_c} u^{11}, \quad 
    u = (\mathcal{M}_c \Omega)^{1/3}.
\end{equation}
Since PN orders scale with powers of $u$ (where $u^{2n}$ corresponds to $n$PN), the $\Lambda^2 u^4$ terms in Eqs.~\eqref{orbitalseparation} and~\eqref{frequencyevolution} confirm the expected 2PN nature of the noncommutative corrections.

\subsection{Gravitational-wave waveform within parameterized post-Einstein framework}
To constrain spacetime noncommutativity (characterized by $\Lambda$) using observed GW events, we employ the ppE framework~\cite{Yunes:2009ke,Cornish:2011ys}. Within this framework, the noncommutative corrections in Eqs.~\eqref{orbitalseparation} and~\eqref{frequencyevolution} correspond to non-GR modifications of ppE GW waveforms~\cite{Tahura:2018zuq}.
The ppE waveform for the $(2,2)$ mode in the Fourier domain, for a compact inspiraling binary system, is given by~\cite{Yunes:2009ke}
\begin{equation}
\tilde{h}(f) = \tilde{h}_{\mathrm{GR}} \left(1 + \alpha u^a \right) e^{i \beta u^b}, \label{ppewaveform}
\end{equation}
where $\tilde{h}_{\mathrm{GR}}$ denotes the GR waveform, while $\alpha u^a$ and $\delta\Psi = \beta u^b$ represent non-GR corrections to the amplitude and phase, respectively. These non-GR effects are fully characterized by the ppE parameters $\{\alpha, \beta, a, b\}$.
Similarly, non-GR modifications to orbital dynamics are expressed in the ppE formalism as~\cite{Kobakhidze:2016cqh}
\begin{align}
    r &= r_{\mathrm{GR}} \left(1 + \gamma_r u^{c_f} \right), \label{orbital2} \\
    \dot{f} &= \dot{f}_{\mathrm{GR}} \left(1 + \gamma_{\dot{f}} u^{c_{\dot{f}}} \right). \label{frequency2}
\end{align}
Since noncommutative corrections to orbital separation and frequency evolution occur at the same PN order in Eqs.~\eqref{orbitalseparation} and~\eqref{frequencyevolution}, we follow Tahura and Yagi~\cite{Tahura:2018zuq} in utilizing relations between ppE parameters $(\alpha,\beta,a,b)$ and orbital parameters $(\gamma_r,c_r,\gamma_{\dot f},c_{\dot{f}})$ for the \textit{comparable dissipative and conservative case}. Comparing Eqs.~\eqref{orbitalseparation},~\eqref{frequencyevolution} with Eqs.~\eqref{orbital2} and~\eqref{frequency2} yields
\begin{align}
    &\alpha_{\mathrm{NC}} = -\frac{3}{8}\eta^{-4/5}(2\eta-1)\Lambda^2,  &a_{\mathrm{NC}} &= 4; \label{alpha} \\
    &\beta_{\mathrm{NC}} = -\frac{75}{256}\eta^{-4/5}(2\eta-1)\Lambda^2, & b_{\mathrm{NC}}&= -1. \label{beta}
\end{align}
We emphasize the spacetime noncommutativity origin of these parameters through the subscript ``$\mathrm{NC}$". Since the current matched-filtering techniques used in terrestrial GW data analysis are more sensitive to the phase than the amplitude \cite{Bonilla:2022dyt}, and the luminosity-distance uncertainties of GW150914 and GW190814 are much larger than the expected ppE amplitude corrections, we safely neglect amplitude corrections in our analysis.

The above deviation shows the leading PN correction of noncommutative gravity to the (2,2) mode of GW in GR. However, recent observations have shown the evidence for the presence of higher harmonics, especially for the GW190814 considered in this work~\cite{LIGOScientific:2020zkf}. The omission of these higher harmonics can lead to a systematic bias for estimating the masses and mass ratio~\cite{Chatziioannou:2019dsz, LIGOScientific:2020zkf}. Thus, Mezzasoma and Yunes~\cite{Mezzasoma:2022pjb} improved the ppE framework by including the $l=2,3$ and $4$ higher harmonics. Therefore, the full inspiral waveform including higher harmonics for noncommutative gravity is,

\begin{equation}
\begin{aligned}
\tilde{h}_{{\rm NC}}(f)&=\sum_{lm}\tilde{h}_{{\rm NC},lm}(f),\\
\tilde{h}_{NC,lm}(f)&=\tilde{h}_{lm}^{GR}(f)\exp[{i\beta_{{\rm NC},lm}u^{b_{{\rm NC},lm}}}],
\end{aligned}\end{equation}
where $\beta_{{\rm NC},lm}$ and $u^{b_{{\rm NC},lm}}$ \cite{Garcia-Quiros:2020qpx, Mehta:2022pcn, Xie:2024xex} are
\begin{equation}
\beta_{{\rm NC},lm}=\left(\frac{2}{m}\right)^{\frac{b_{{\rm NC}}}{3}-1}\beta_{{\rm NC}},\quad u^{b_{{\rm NC},lm}}=(\frac{2}{m}\mathcal{M}_c \pi f)^{b_{\rm NC}}.
\end{equation}

Moreover, the full GW waveform of CBC event also includes the merger and ringdown phase, which contributes to the most partition of SNR. Bonilla {\it et al.}~\cite{Bonilla:2022dyt} proposed a new extension to the ppE framework for allowing the corrections to GR waveform past the inspiral phase. In this work, we consider the $C^0$ correction for merger-ringdown phase~\cite{Bonilla:2022dyt},

\begin{equation}
\tilde{h}_{\ell m}^{\mathcal{C}^0}(f)=
    \begin{cases}
        \tilde{h}_{\ell m}^{\mathrm{GR}}e^{i\beta_{\ell m}u^{b_{\ell m}}},&f<f_{\ell m}^{\mathrm{IM}}\\
        \tilde{h}_{\ell m}^{\mathrm{GR}}e^{i\beta_{\ell m}u_{\mathrm{IM}}^{b_{\ell m}}},&f\geq f_{\ell m}^{\mathrm{IM}}
    \end{cases},
\end{equation}
where the $f_{lm}^{\rm IM}=mf_{\mathrm{~MECO}}^{22}$/2 is the transition frequency for each mode and $f_{\mathrm{~MECO}}^{22}= 0.018/M$ is the minimal energy circular orbit (MECO) frequency for (2,2) mode in detector frame for the \texttt{IMRPhenomXHM} template~\cite{Garcia-Quiros:2020qpx}. 

Thus, in this work, we utilize the noncommutative gravity waveform $\tilde{h}_{\ell m}^{\mathcal{C}^0}$ which includes the full inspiral, merger, and ringdown phases, and both the dominant (2,2) mode and several higher-order modes, including (2,1), (3,3), (3,2), and (4,4). As show in \myfig{waveforms}, we plot the real part of $h_\times$ for a GW150914-like event under both GR and noncommutative gravity. In this work, we set the low-frequency cutoff and the reference frequency to 20 Hz. The left panel represents the noncommutative waveform corresponding to the upper bound on $\sqrt{\Lambda}$ obtained in Ref.~\cite{Kobakhidze:2016cqh} and the 95th percentile upper bound  obtained in this work. The right panel presents a zoomed-in view of the waveform in the frequency range from 20 Hz to 22 Hz. Notably, as shown in Eq.~\eqref{beta}, the phase correction is positive for $\eta \in (0, 0.25]$. Following the convention in Yunes and Pretorius~\cite{Yunes:2009ke}, as well as Tahura and Yagi~\cite{Tahura:2018zuq}, the GR waveform is expressed as $\tilde{h}^{\mathrm{GR}}=A_{\rm GR}e^{i\Psi_{\rm GR}}$. Under this convention, a positive non-GR phase correction leads to an overall leftward shift of the waveform. The non-GR waveform employed in this work exhibits behavior consistent with this expectation, as illustrated in the right panel of \myfig{waveforms}. 


\begin{table*}[ht]
\caption{The table of prior distributions for the GR and noncommutative gravity for GW150914 and GW190814.\label{priors}}
\begin{tabular}{lcccc}
\toprule
Parameters & \ GW150914 (GR) \ & \ GW150914 (NC) \ & \ GW190814 (GR) \ & \ GW190814 (NC) \ \\
\midrule
$\mathcal{M}_c$& $\mathcal{U}[23, 42]~{\rm M_{\odot}}$ & $\mathcal{U}[23, 42]~{\rm M_{\odot}}$ & $\mathcal{U}[5, 7]~{\rm M_{\odot}}$ & $\mathcal{U}[5, 7]~{\rm M_{\odot}}$ \\
$q$& $\mathcal{U}[1, 4]$ & $\mathcal{U}[1, 4]$ & $\mathcal{U}[2, 20]$ & $\mathcal{U}[2, 20]$ \\
$\chi_{1z}$& $\mathcal{U}[-1, 1]$ & $\mathcal{U}[-1, 1]$ &  $\mathcal{U}[-1, 1]$ & $\mathcal{U}[-1, 1]$ \\
$\chi_{2z}$& $\mathcal{U}[-1, 1]$ & $\mathcal{U}[-1, 1]$ & $\mathcal{U}[-1, 1]$ & $\mathcal{U}[-1, 1]$ \\
$\sqrt\Lambda$& -- & $\mathcal{U}[0, 20]$ & -- & $\mathcal{U}[0, 20]$ \\
$D_L$& $\mathcal{U}[200, 800]~{\rm Mpc}$ & $\mathcal{U}[200, 800]~{\rm Mpc}$ & $\mathcal{U}[5, 500]~{\rm Mpc}$ & $\mathcal{U}[5, 500]~{\rm Mpc}$ \\
$\iota$& sine uniform &  sine uniform  & sine uniform  & sine uniform  \\
$\alpha$& uniform\ sky & uniform\  sky & uniform\  sky & uniform\  sky \\
$\delta$& uniform\ sky & uniform\  sky & uniform\  sky & uniform\  sky \\
$\psi$& $\mathcal{U}[0, 2\pi]$ & $\mathcal{U}[0, 2\pi]$ & $\mathcal{U}[0, 2\pi]$ & $\mathcal{U}[0, 2\pi]$ \\
$\delta t_c$& $\mathcal{U}[-0.1, 0.05]~\mathrm{s}$ & $\mathcal{U}[-0.1, 0.05]~\mathrm{s}$ &  $\mathcal{U}[-0.1, 0.05]~\mathrm{s}$ &  $\mathcal{U}[-0.1, 0.05]~\mathrm{s}$ \\
$\phi_c$& $\mathcal{U}[0, 2\pi]$ & $\mathcal{U}[0, 2\pi]$ & $\mathcal{U}[0, 2\pi]$ &  $\mathcal{U}[0, 2\pi]$\\
\bottomrule
\end{tabular}
\end{table*}

\section{Data analysis\label{data_analysis}}
This section begins with an overview of GW150914 and GW190814, highlighting the motivation for selecting these events to constrain noncommutative gravity. We then describe the Bayesian inference methodology employed in this analysis.

\subsection{GW150914}
On September 14, 2015, the first GW event, dubbed GW150914, was observed by the LIGO Livinston and LIGO Hanford detectors. This signal was generated by the coalescence of a BBH system with component masses of ${36~\rm M}_{\odot}$ and ${29~\rm M}_{\odot}$~\cite{LIGOScientific:2020zkf}. With a network SNR of 24 and a false alarm rate of less than one event per 203,000~years, GW150914 offered a unique opportunity to test GR in the strong-field, highly dynamical regime. Detailed analyses have been performed on this event in a theory-independent manner, and no evidence for deviations from GR have been founded~\cite{LIGOScientific:2016lio}.

Kobakhidze {\it et al.}~\cite{Kobakhidze:2016cqh} constrained noncommutative gravity by analyzing the phase deviation from GR at the 2PN order~\cite{LIGOScientific:2016lio}, as measured in GW150914, and obtained an upper bound of $\sqrt{\Lambda}<3.5$. In this work, we reanalyze GW150914 using the ppE formalism within a Bayesian framework. We utilize the observed data of GW150914 from LIGO Hanford and LIGO Livingston, publicly available through the Gravitational Wave Open Science Center~(GWOSC)~\cite{GWOSC_GWTC}. We performed the analysis on observed data spanning from 20 seconds before to 10 seconds after the trigger time \cite{Nitz:2021zwj}, with both the low-frequency cutoff and reference frequency set to 20~Hz.

\subsection{GW190814}
On August 14, 2019, GW190814 was observed during the third observation run of LIGO and Virgo~\cite{LIGOScientific:2020zkf}. With a network SNR of approximately 25, the event corresponds to the coalescence of a binary system consisting of ${23~\rm M}_{\odot}$ BH and a ${2.6~\rm M}_{\odot}$ compact companion. The mass ratio of this event is 8.9, making it the most asymmetric binary system among all events reported in GWTC-1, GWTC-2, GWTC-2.1, and GWTC-3~\cite{GWOSC_GWTC}. The corresponding symmetric mass ratio is $\eta \sim 0.09$, further highlighting the extreme mass asymmetry of the system.

\begin{figure}[ht]
  \centering
  \includegraphics[width=0.4\textwidth]{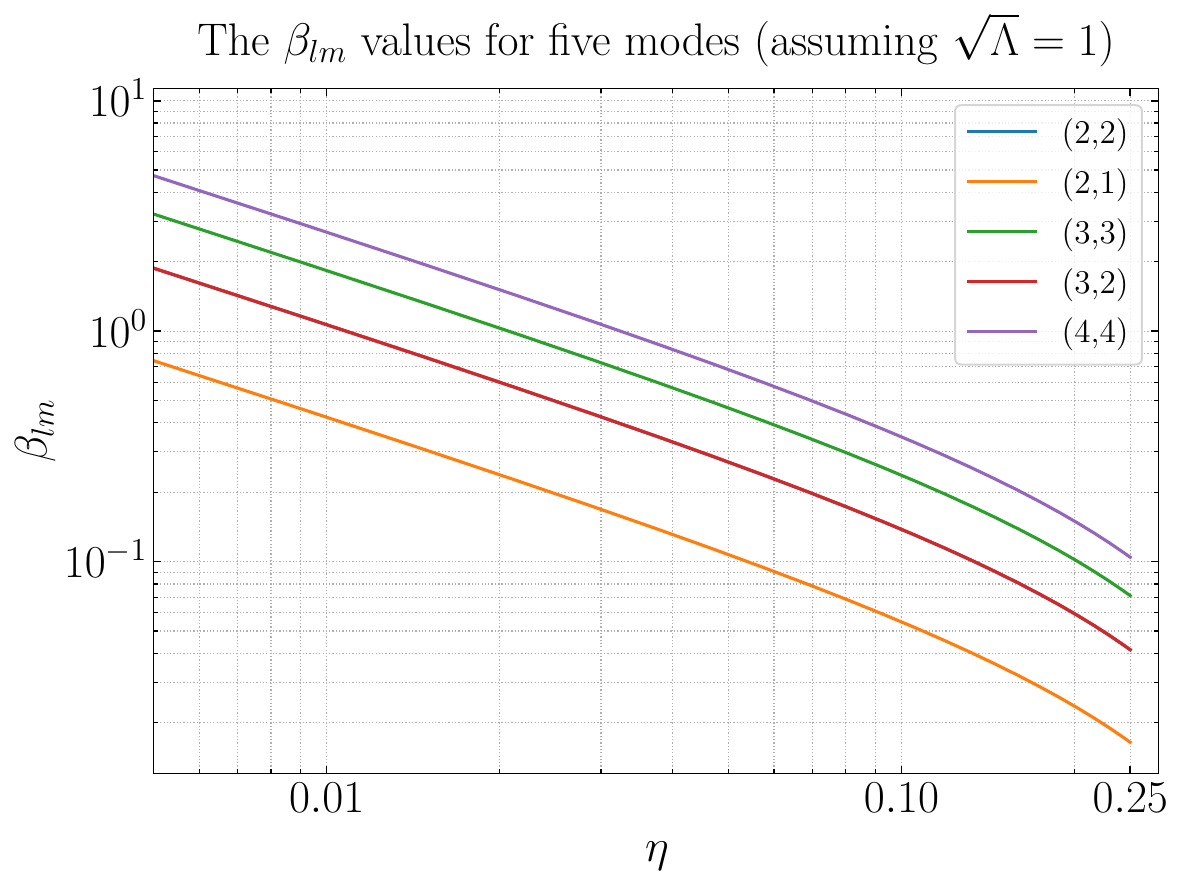}
  \caption{The values of the noncommutative gravity phase correction $\beta_{lm}$ are presented for five modes: (2,2), (2,1), (3,3), (3,2), and (4,4), assuming $\sqrt{\Lambda} = 1$. The $\beta_{lm}$ value for the (2,2) mode is degenerate  with that of the (3,2) mode according to Eq.~\eqref{beta}.
  \label{fig_beta_lm}}
\end{figure}

As shown in \myfig{fig_beta_lm}, the value of $\beta_{lm}$ is inversely proportional to the symmetric mass ratio $\eta$. Modes with larger $m$ exhibit higher values of $\beta_{lm}$. However, for the GR contribution, the (2,2) mode dominates, and consequently, the full waveform in noncommutative gravity is also dominated by the (2,2) mode. Meanwhile, for all modes, the more extreme the symmetric mass ratio $\eta$ is, the higher the value of $\beta_{lm}$ is. This is the main reason we select the event GW190814, which has the most extreme symmetric mass ratio and a relatively high SNR. 
In this work, we utilize the observed data of GW190814 from LIGO Hanford and LIGO Livingston detectors, obtained from the GWOSC~\cite{GWOSC_GWTC}. Treating the source as a BBH system, we performed the analysis on observed data spanning from 20 seconds before to 10 seconds after the trigger time \cite{Nitz:2021zwj}, with both the low-frequency cutoff and reference frequency set to 20 Hz.

\subsection{Bayesian inference}

In this work, we constrain both GR and non-GR parameters from the observed GW events in the Bayesian framework~\cite{Finn:1992wt, Lyu:2022gdr}. In this approach, the posterior distribution of parameters $p({\vartheta}|d,\mathcal{H})$, given the observed data $d$ and hypothesis $\mathcal{H}$ is given by,
\begin{equation}
p(\vartheta|d,\mathcal{H})=\frac{p(d|\vartheta,\mathcal{H})\mathrm{~}p(\vartheta|\mathcal{H})}{p(d|\mathcal{H})},
\end{equation}
where $p(d|\vartheta,\mathcal{H})$ denotes the likelihood, $p(\vartheta|\mathcal{H})$ is the prior, and $p(d|\mathcal{H})$ presents the evidence. For a network of multiple detectors, the likelihood is given by,
\begin{equation}
\log p(d|\vartheta,\mathcal{H})=\log\bar{\alpha}-\frac{1}{2}\sum_i\left\langle d_i-h_i(\vartheta)|d_i-h_i(\vartheta)\right\rangle,
\end{equation} 
where $\log\bar{\alpha}$ is the normalizing constant, the subscript $i$ denotes the $i$th detector, and $h$ presents the waveform template. The inner product is defined as, 
\begin{equation}
    \langle a(t)|b(t)\rangle=2\int_{f_{\mathrm{low}}}^{f_{\mathrm{high}}}\frac{\tilde{a}^*(f)\tilde{b}(f)+\tilde{a}(f)\tilde{b}^*(f)}{S_n(f)}df.
\end{equation}
where $f_{\mathrm{low}}$ and $f_{\mathrm{high}}$ are the low- and high-frequency cutoffs of the data, respectively. 
Since both GW150914 \cite{LIGOScientific:2016wkq} and GW190814 \cite{LIGOScientific:2020zkf} exhibit negligible evidence for spin precession, and it has been shown that precessing and nonprecessing waveform models yield consistent posterior distributions for GW150914 \cite{LIGOScientific:2016wkq}, we model both events as spin-aligned binary black hole systems in this work. Thus, the GR waveform can be described with 11 parameters, 
\begin{equation}
\vartheta_{\rm GR}=\{\mathcal{M}_c,q,\chi_{1z},\chi_{2z},D_L,\iota, \alpha,\delta,\psi,t_c,\phi_c\},
\end{equation}
where $\mathcal{M}_c$ and $q$ are chirp mass and the mass ratio,
$\chi_{1z}$ and $\chi_{2z}$ are their dimensionless spin components which
parallel with the orbital angular momentum, $D_L$ is the luminosity distance of the source, $\alpha$ and $\delta$ are the right ascension and declination angles respectively, $\iota$ and $\psi$ are the inclination angle the polarization angle respectively, and $\phi_c$ and $t_c$ are the coalescence phase and time respectively. 

For the waveform of noncommutative gravity, additional non-GR parameter $\sqrt\Lambda$ will be considered,
\begin{equation}
\vartheta_{\rm NC}=\{\mathcal{M}_c,q,\chi_{1z},\chi_{2z},\sqrt\Lambda, D_L,\iota, \alpha,\delta,\psi,t_c,\phi_c\}.
\end{equation}
The prior distributions of all parameters are summarized in \mytab{priors}. In this work, we perform Bayesian inference using the \textsc{PyCBC} package~\cite{Biwer:2018osg}.

\begin{figure*}[ht] 
	\centering 
    \begin{minipage}{0.46\textwidth}
        \centering
        \includegraphics[width=0.95\linewidth]{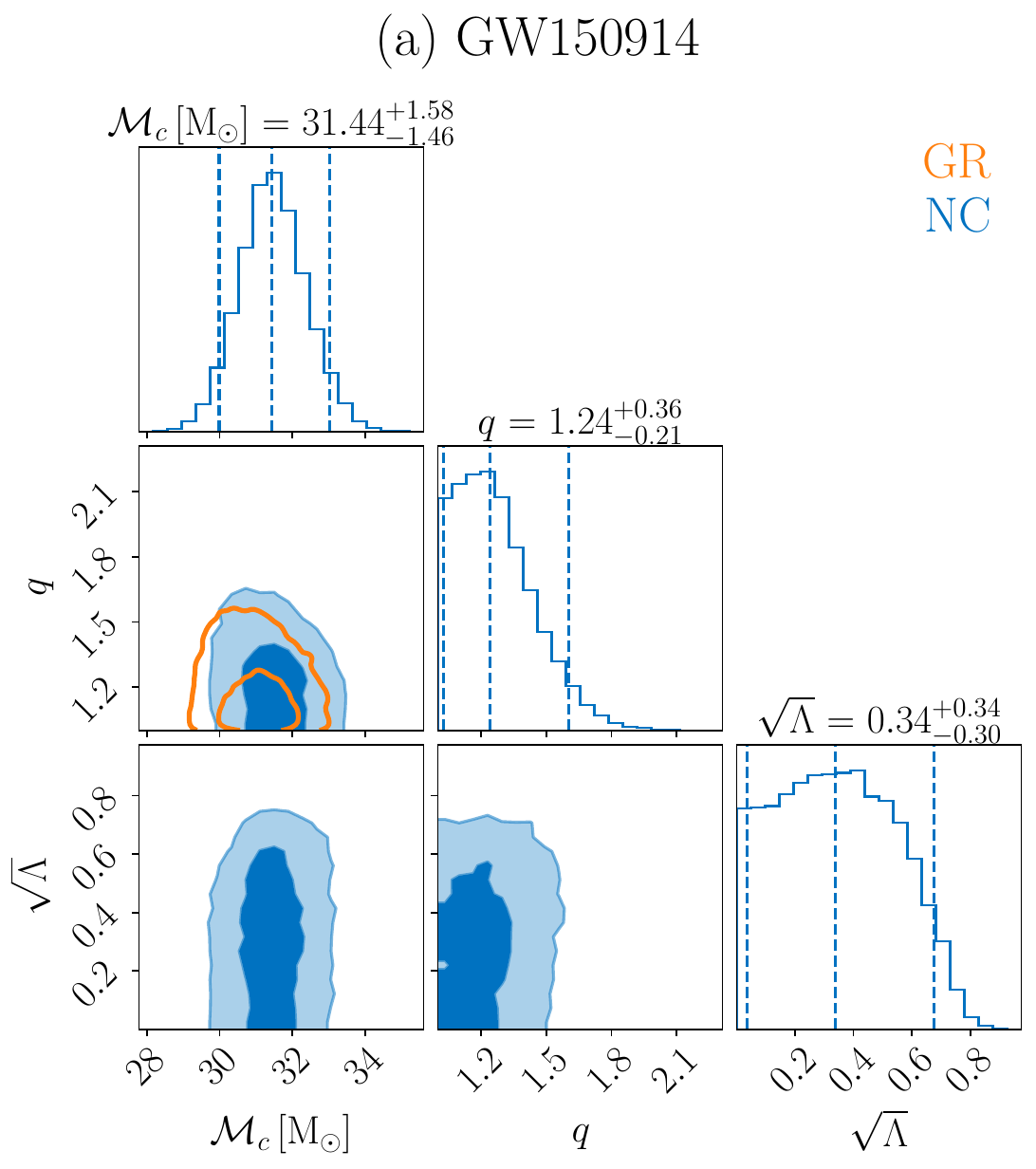}
    \end{minipage}\hfill
    \begin{minipage}{0.47\textwidth}
        \centering
        \includegraphics[width=0.95\linewidth]{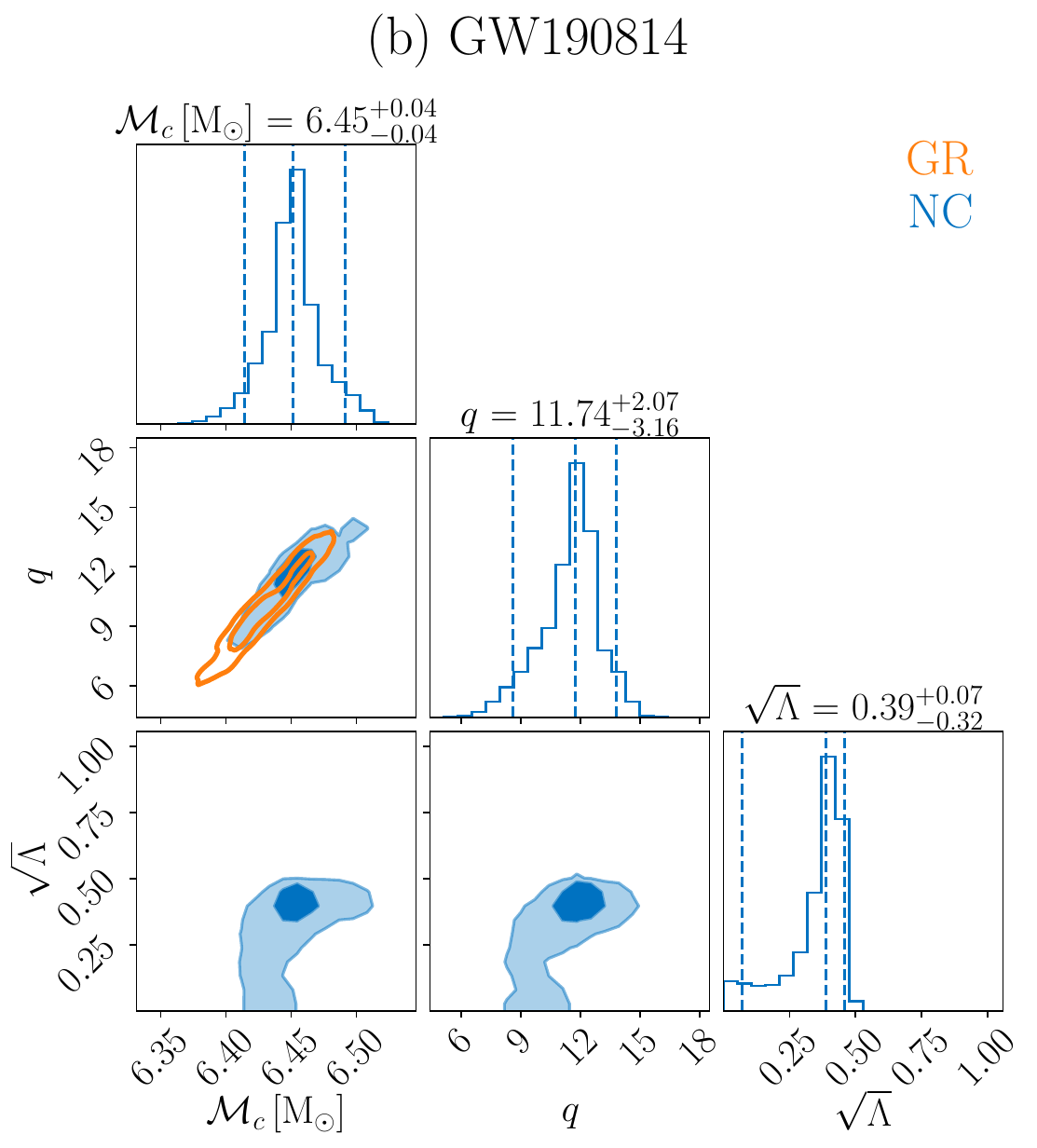} 
    \end{minipage}
    \caption{Posterior distributions for GW150914 (left panel) and GW190814 (right panel). The GR results are shown as orange contours, while the results for noncommutative gravity are depicted in blue. The 2D contours represent with different credible intervals, denoting the 50\% and 90\% regions, while the vertical lines indicate the  90\% region for the 1D marginalized posterior distribution.
    \label{results}}
\end{figure*}

\section{Results and discussion\label{sec:results}}
In this work, we analyze two GW events, GW150914 and GW190814, to constrain noncommutative gravity within the ppE framework. 

We first analyze the two events within the framework of GR, using the \texttt{IMRPhenomXHM} waveform model. The orange contours in \myfig{results} show the GR posterior distributions for key parameters, including the chirp mass $\mathcal{M}_c$ and mass ratio $q$. The 90\% credible interval for $\mathcal{M}_c$ and $q$ are $31^{+1.5}_{-1.5}~{\rm M_{\odot}}$ and $1.15^{+0.37}_{-0.14}$ for GW150914, and $6.43^{+0.046}_{-0.034}~{\rm M_{\odot}}$ and $9.9^{+3.2}_{-2.9}$ for GW190814, respectively. These results, for both the primary and the remaining nine parameters, are consistent with those reported in Ref.~\cite{Nitz:2021zwj}. 

Next, we employ the noncommutative gravity waveform constructed within the ppE framework to constrain the non-GR parameter $\sqrt{\Lambda}$ with GW150914 and GW190814. The same Bayesian inference settings as in the GR case are adopted, except for the waveform template and the prior distribution of  $\sqrt{\Lambda}$, as summarized in \mytab{priors}. The posterior distributions of noncommutative gravity are shown in blue in \myfig{results}. The inferred GR parameters such as $\mathcal{M}_c$ and $q$ are consistent with those obtained in the GR-only analysis of the same GW events. For example, the 90\% credible intervals for $\mathcal{M}_c$ and $q$ are $31.44^{+1.58}_{-1.46}~{\rm M_{\odot}}$ and $1.24^{+0.36}_{-0.21}$ for GW150914, and $6.45^{+0.04}_{-0.04}~{\rm M_{\odot}}$ and $11.74^{+2.07}_{-3.16}$ for GW190814, respectively. The remaining nine GR parameters inferred  within the noncommutative gravity framework are also consistent with those obtained under GR. These findings further confirm the robustness and consistency of the inferred noncommutative gravity parameter. 

For the non-GR parameter $\sqrt{\Lambda}$, the 90\% credible intervals are $0.34^{+0.34}_{-0.30}$ for GW150914 and $0.39^{+0.07}_{-0.32}$ for GW190814. The corresponding 95th percentile upper bounds are 0.68 and 0.46, respectively. The tighter constraint on $\sqrt{\Lambda}$ obtained from GW190814 than GW150914 is
also consistent with the phase correction in Eq.~\eqref{beta}, since the GW190814 has a more extreme symmetric mass ratio. 
It is therefore useful to elucidate the physical implications of this constraint. For convenience, we henceforth set \(c=1\).
Recall the definition of \(\Lambda\),
\begin{equation}
\Lambda \theta^i \equiv \frac{\theta^{0i}}{l_P t_P} = \theta^{0i} E_P^2,
\end{equation}
where \(\vec{\theta} = (\theta^1,\theta^2,\theta^3)\) is a unit vector specifying the direction of \(\theta^{0i}\) in space and \(E_P \simeq 1.2 \times 10^{19}\,\mathrm{GeV}\) denotes the Planck energy. This immediately implies
\begin{equation}
\Lambda = \lVert \theta^{0i}\rVert\times  E_P^2 .
\end{equation}
Using the constraint obtained in this work, the characteristic energy scale at which spacetime noncommutativity becomes relevant is bounded from below as
\begin{equation}
\frac{1}{\sqrt{\lVert \theta^{0i} \rVert}} > \frac{E_P}{0.46} \simeq 2.2\, E_P \simeq 2.7 \times 10^{19}~\mathrm{GeV}.
\end{equation}
Equivalently, this may be interpreted as an upper bound on the noncommutative~(NC) length scale,
\begin{equation}
\sqrt{\lVert \theta^{0i} \rVert} <  0.46\, l_P \simeq 0.7 \times 10^{-35}~\mathrm{m}.
\end{equation}
We therefore find that the NC length scale is mildly smaller than the Planck length \(l_P\), indicating that noncommutativity between space and time is extremely suppressed. Moreover, it is natural to expect that the space--space components of \(\theta^{\mu\nu}\) are at least of the same order of magnitude as the space--time components \(\theta^{0i}\).~\footnote{Since the time--space components might violate the unitarity of S--matrices, they are naturally expected to be much smaller than the space--space components~\cite{Gomis:2000zz}.}
Under this assumption, our result implies that observable effects of spacetime noncommutativity are negligible except at energies approaching the Planck scale.
This bound is significantly stronger than the theoretical estimate
\(1/\sqrt{\lVert \theta^{0i} \rVert} \gtrsim 10~\mathrm{TeV}\)
derived in Ref.~\cite{Calmet:2004dn}, as well as those~(\(1/\sqrt{\lVert\theta^{0i}\rVert}\gtrsim \mathcal{O}(10)~\text{TeV}\) as well) obtained from various phenomenological analyses~\cite{Chaichian:2000si,Carroll:2001ws,Joby:2014oee,Calmet:2015fma}. It is also manifestly stronger than the constraint
\(1/\sqrt{\lVert \theta^{0i} \rVert} > 5 \times 10^{14}~\mathrm{GeV}\)
reported in Ref.~\cite{Mocioiu:2000ip}, exceeding it by nearly five orders of magnitude. Our result is instead comparable to the bound obtained in Ref.~\cite{Kobakhidze:2016cqh}.
Although such constraints are inevitably model-dependent, bounds on spacetime noncommutativity are nonetheless expected to lie within a similar range. In this regard, it is evident that GW observations provide the most stringent constraints currently available on spacetime noncommutativity.

Previous constraints on noncommutative gravity from GW observations in Kobakhidze {\it et al.} \cite{Kobakhidze:2016cqh} and Jenks {\it et al.} \cite{Jenks:2020gbt} were mainly derived from projection-based analyses of the LIGO-Virgo tests of general relativity~\cite{LIGOScientific:2016lio, LIGOScientific:2019fpa, LIGOScientific:2018dkp}, in which deviations from GR are parametrized by PN coefficients and logarithmic coefficients. The 2PN deviation $\delta\hat{\varphi}_4$ was mapped onto the noncommutative gravity parameter $\sqrt{\Lambda}$ \cite{Kobakhidze:2016cqh,Jenks:2020gbt}. These studies relied on inspiral-only information and yielded bounds at $\sqrt{\Lambda}\lesssim3.5$ and $\sqrt{\Lambda}\lesssim1.5$ for Kobakhidze {\it et al.} \cite{Kobakhidze:2016cqh} and Jenks {\it et al.} \cite{Jenks:2020gbt}, respectively. While the tighter constraint obtained in this work, $\sqrt{\Lambda}<0.46$, results from several complementary improvements. First, we perform a full Bayesian inference, directly constraining the noncommutative parameter from the GW data while consistently accounting for waveform systematics and parameter correlations, in contrast to projection-based estimates. Second, by including the $C^0$-corrected ppE waveform \cite{Bonilla:2022dyt}, we are able to exploit the full inspiral–merger–ringdown signal rather than restricting to the inspiral stage. Third, the noncommutative effect is enhanced for systems with extreme symmetric mass ratios $\eta$, motivating our analysis of GW190814, which provides substantially stronger sensitivity than more symmetric binaries. Finally, the use of the \texttt{IMRPhenomXHM} waveform model, including higher-order modes, is crucial for accurately modeling such asymmetric systems and further improves the constraint.

In this work, we consider the aligned-spin BBH events and the leading PN correction of noncommutative gravity within the ppE framework. Although explicit spin–orbit and spin–spin terms are not included in the construction of the noncommutative gravity waveform derived in Kobakhidze {\it et al.}, some spin effects can nevertheless enter the 2PN ppE correction through precession-induced modulations, as discussed in Jenks {\it et al.} This therefore introduces additional degeneracies between $\sqrt{\Lambda}$ and the spin parameters, which are expected to moderately weaken the constraint on $\sqrt{\Lambda}$, as illustrated in previous studies of spin-induced parameter degeneracies \cite{Baird:2012cu, Song:2024pnk}. A systematic investigation of these effects would be pursued in future. Moreover, with the deployment of next-generation terrestrial GW observatories such as Cosmic Explorer~\cite{Reitze:2019iox}, Einstein telescope~\cite{Punturo:2010zz}, and space-borne detectors including LISA~\cite{LISA:2017pwj}, Taiji~\cite{Hu:2017mde, Du:2025xdq} and TianQin~\cite{TianQin:2015yph}, as well as the Lunar Gravitational-wave
Antenna~(LGWA)~\cite{LGWA:2020mma}, a wider variety of interesting systems with higher SNR and more extreme symmetric mass ratio will be detected. These include the intermediate-mass black hole binaries~\cite{Reali:2024hqf, Song:2025lpa, Dong:2025ikq} and extreme mass-ratio inspiral~\cite{Babak:2017tow, Lyu:2024gnk, Liang:2024qzf}. Such detections will provide valuable opportunities for probing noncommutative gravity in the future.

In summary, this work presents a comprehensive analysis constraining noncommutative gravity, yielding a 95th percentile upper bound of $\sqrt{\Lambda}<0.46$ for GW190814, representing the strongest limit on noncommutative gravity derived from real GW data so far, and corresponding to a characteristic energy scale above $2.2\,E_P$ or a length scale below $0.46\,l_P$.

\section*{Acknowledgements}
%
We thanks Haobo Yan, Yiqi Xie, and the anonymous referee for profound comments.
This work is supported in part by the National Science Foundation of China under Grant No.~12547101.
H.L. is also supported by the start-up fund of Chongqing University under No.~0233005203009, Z.L. is supported by ``the Fundamental Research Funds for the Central Universities'', and J.Z. is supported by the start-up fund of Chongqing University under No.~0233005203006. This work is also supported by the High-performance Computing Platform of Peking University.

\bibliography{scibib}

\end{document}